\documentclass[aps,prl,reprint,twocolumn,amsmath,amssymb]{revtex4-1}

\pdfoutput=1

\usepackage{graphicx}
\usepackage{dcolumn}
\usepackage{bm}

\newcommand{\mr}[1]{\mathrm{#1}}
\newcommand{\be}{\begin{equation}}
\newcommand{\ee}{\end{equation}}

\newcommand{\figta}{$\left(\mathrm{a}\right)\;$}
\newcommand{\figtb}{$\left(\mathrm{b}\right)\;$}

\newcommand{\figa}{$\left(\mathrm{a}\right)$}
\newcommand{\figb}{$\left(\mathrm{b}\right)$}

\newcommand{\Mohm}{\;\mr{M}\Omega}

\newcommand{\hz}{\;\mr{Hz}}
\newcommand{\khz}{\;\mr{kHz}}
\newcommand{\mhz}{\;\mr{MHz}}

\newcommand{\mk}{\;\mr{mK}}

\newcommand{\s}{\;\mr{s}}

\newcommand{\ns}{\;\mr{ns}}
\newcommand{\ms}{\;\mr{ms}}

\newcommand{\mv}{\;\mr{mV}}

\newcommand{\mum}{\;\mu\mr{m}}

\newcommand{\mev}{\;\mr{meV}}
\newcommand{\muv}{\;\mu\mr{V}}

\newcommand{\pa}{\;\mr{pA}}
\newcommand{\muev}{\;\mu\mr{eV}}
\newcommand{\nm}{\;\mr{nm}}

\newcommand{\vdut}{V_{\mr{b}}}
\newcommand{\idut}{I}
\newcommand{\vdet}{V_{\mr{det}}}
\newcommand{\vgdet}{V_{\mr{g,det}}}
\newcommand{\vgl}{V_{\mr{g,L}}}
\newcommand{\vgr}{V_{\mr{g,R}}}
\newcommand{\vgm}{V_{\mr{g,M}}}
\newcommand{\esigma}{E_{\mr{c},\Sigma}}
\newcommand{\idet}{I_{\mr{det}}}
\newcommand{\ag}{A_{\mr{g}}}

\newcommand{\rtot}{R_{\Sigma}}
\newcommand{\rtdet}{R_{\mr{T,det}}}
\newcommand{\frep}{f}
\newcommand{\fs}{f_{\mr{S}}}
\newcommand{\trep}{T_{\mr{rep}}}
\newcommand{\tpulse}{T_{\mr{pulse}}}
\newcommand{\taudet}{\tau_{\mr{det}}}

\newcommand{\gammaeff}{\Gamma_{\mr{eff}}}

\begin{document}

\title{On-chip error counting for hybrid metallic single-electron turnstiles}

\author{J. T. Peltonen}
\email{joonas.peltonen@aalto.fi}
\affiliation{Low Temperature Laboratory, Department of Applied Physics, Aalto University School of Science, POB 13500, FI-00076 AALTO, Finland}

\author{V. F. Maisi}
\altaffiliation[Present Address: ]{ETH Zurich, Switzerland}
\affiliation{Low Temperature Laboratory, Department of Applied Physics, Aalto University School of Science, POB 13500, FI-00076 AALTO, Finland}

\author{S. Singh}
\affiliation{Low Temperature Laboratory, Department of Applied Physics, Aalto University School of Science, POB 13500, FI-00076 AALTO, Finland}

\author{E. Mannila}
\affiliation{Low Temperature Laboratory, Department of Applied Physics, Aalto University School of Science, POB 13500, FI-00076 AALTO, Finland}

\author{J. P. Pekola}
\affiliation{Low Temperature Laboratory, Department of Applied Physics, Aalto University School of Science, POB 13500, FI-00076 AALTO, Finland}

\date{\today}

\begin{abstract}
We perform in-situ detection of individual electrons pumped through a single-electron turnstile based on ultrasmall normal metal -- insulator -- superconductor tunnel junctions. In our setup, limited by the detector bandwidth, at low repetition rates we observe errorless sequential transfer of up to several hundred electrons through the system. At faster pumping speeds up to $100\khz$, we show relative error rates down to $10^{-3}$, comparable to typical values obtained from measurements of average pumped current in non-optimized individual turnstiles. The work constitutes an initial step towards a self-referenced current standard realized with metallic single-electron turnstiles, complementing approaches based on semiconductor quantum dot pumps. It is the first demonstration of on-chip pumping error detection at operation frequencies exceeding the detector bandwidth, in a configuration where the average pumped current can be simultaneously measured. The scheme in which electrons are counted from the superconducting lead of the turnstile, instead of direct probing of the normal metal island, also enables studies of fundamental higher-order tunneling processes in the hybrid structures, previously not in reach with simpler configurations.
\end{abstract}

\maketitle

\section{Introduction}
The SI base unit for electric current, the ampere, awaits redefinition in terms of the quantized current $I=nef$ produced by a single-electron pump, transferring an integer number $n$ of electrons during each cycle of the periodic drive at operating frequency $f$~\cite{pekola13}. Recently, a quantum dot-based electron pump~\cite{blumenthal07} combined with ultrastable and sensitive amplifiers for measuring the average current~\cite{drung15} was reported for the first time to surpass the present SI realization of the ampere, reaching relative uncertainty below 0.2 ppm~\cite{stein15}.

For a quantum current standard to meet the stringent goals of a current of approximately $100\pa$ and relative error rates in the $10^{-8}$ range requires the development of the single-charge pumps themselves as well as the current instrumentation. A complementary approach~\cite{wulf08,wulf13} to improve the accuracy is to reliably detect on the level of individual electrons, with an on-chip electrometer such as a radio-frequency SET (single-electron transistor)~\cite{schoelkopf98} or quantum point point contact~\cite{reilly07}, both the number and direction of small deviations from the expected number $nf$ of charges pumped per second. Such self-referenced operation of semiconductor quantum dot pumps has recently been investigated at low repetition rates~\cite{fricke14,fricke13,tanttu15}. In this work we implement a rudimentary version of the scheme considered theoretically in Refs.~\onlinecite{wulf08,wulf13}. We demonstrate its feasibility for metallic SINIS turnstile devices~\cite{pekola08,averin08,pekola13} -- a SET with two superconducting electrodes connected to a normal metal island via two tunnel junctions, functioning as a turnstile for single electrons due to the superconducting energy gap in the leads.

\begin{figure}[!htb]
\includegraphics[width=\columnwidth]{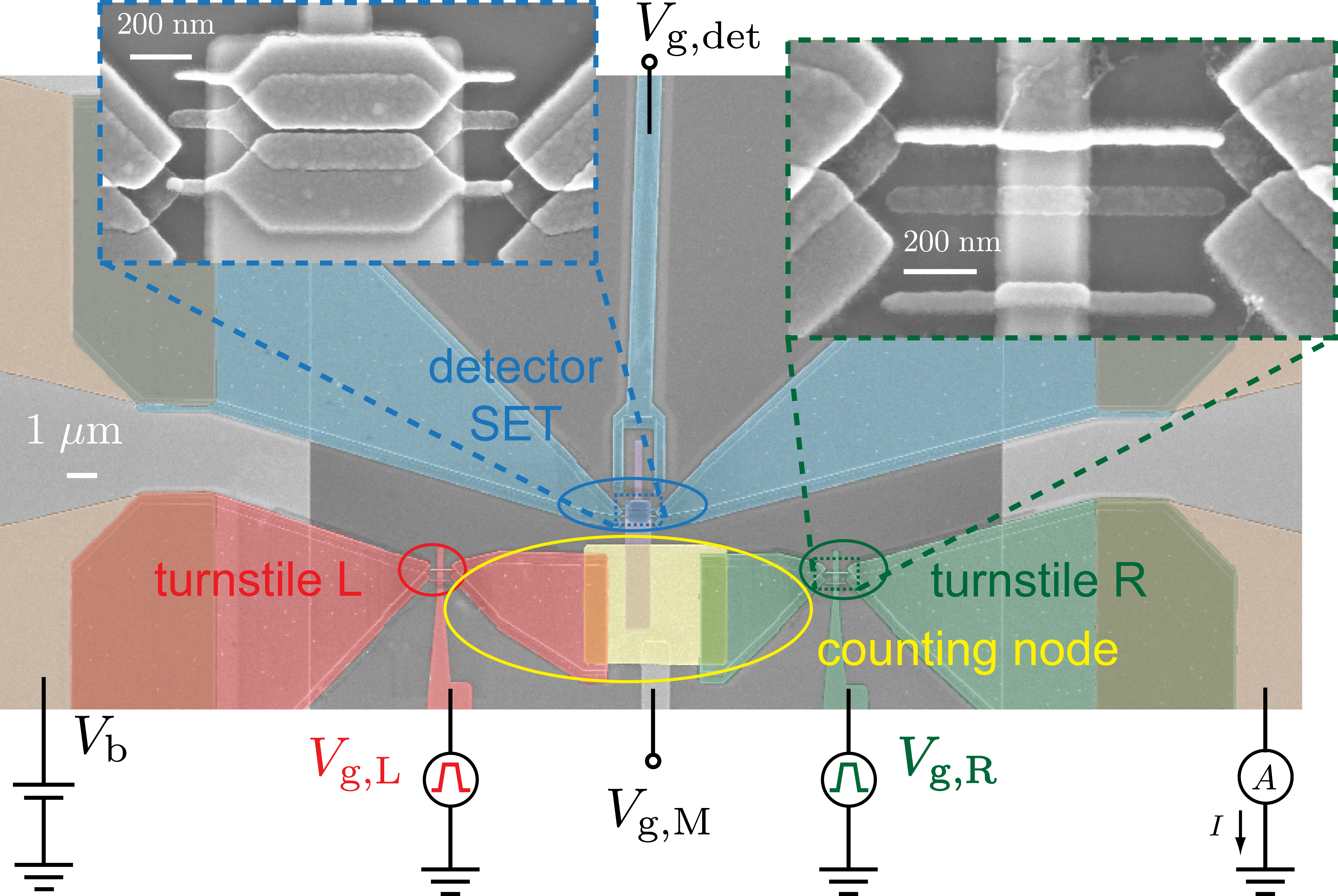}
\caption{(color online) Scanning electron micrograph of an electron counting sample, together with a partial sketch of the measurement scheme. Two SINIS turnstiles [left (L, highlighted in red), and right (R, green)] are connected in series, and a third SINIS (detector, blue), functioning as an electrometer, is capacitively coupled to the metallic node between the L and R turnstiles. The detector SET and one of the turnstiles are shown in enlarged views.} \label{fig:scheme}
\end{figure}

The system, illustrated by the false-color scanning electron micrograph in Fig.~\ref{fig:scheme}, consists of two SINIS turnstiles connected in series. The charge state on finite-sized island or `counting node' between them is monitored in real time by a capacitively coupled single-electron transistor. At low pumping frequencies $\frep=\trep^{-1}\lesssim 200\hz$ and when the turnstiles are sequentially driven, we are able to detect each electron passing the middle island.

Importantly, beyond the detection of each passing charge, at higher $\frep$ when the detector can no longer resolve the individual tunneling events, we show that the rarely occurring pumping errors can nevertheless be readily counted, thus taking the first step towards metallic turnstile operation with error accounting. More generally, our work serves as the first demonstration of pumping error detection at operation frequencies exceeding the detector bandwidth, in a configuration that allows simultaneous measurement of the average pumped current: Very accurate error detection with a slow detector is routinely performed with metallic fully normal-state electron pumps~\cite{pothier92} but only in a $\pm 1e$ shuttling mode~\cite{martinis94,keller96}. On the other hand, with semiconductor quantum dot pumps where directional error counting has already been successfully realized at low repetition rates~\cite{fricke14}, operation at higher duty cycles has not been reported yet.

\section{Sample fabrication process and measurement setup}
The sample in Fig.~\ref{fig:scheme} was fabricated using a process involving three steps of electron beam lithography (EBL) and electron beam evaporation of the thin metal films. Starting with a silicon substrate covered by $300\nm$ thick layer of thermally grown silicon oxide, a pattern was first exposed for a continuous ground plane electrode [light gray in Fig.~\ref{fig:scheme}, starting approximately $5\mum$ away from the NIS junctions] for the purpose of further on-chip filtering of spurious microwave photons reaching the sample~\cite{pekola10}. At the same time, patterns were defined for the gate electrodes as well as a floating coupler electrode (colored light purple) between the detector island and the counting node. These structures were metallized with $2\nm$ of Ti and $30\nm$ of Au, again covered by $2\nm$ of Ti. A $50\nm$ thick $\mr{Al}_2\mr{O}_3$ dielectric layer was then grown on top by atomic layer deposition (ALD) to isolate them from the bias leads and tunnel junction structures to be defined in later steps.

The turnstile gate electrodes were realized as parallel plate capacitors for enhanced gate coupling -- see the enlarged view of turnstile R where the galvanically isolated gate lead is visible as the vertical stripe under the narrow horizontal N island. Moreover, for larger detector signal-to-noise ratio, we further increase the coupling capacitance between the detector and the counting node by creating a central bulge on the detector island to have maximum overlap with the floating electrode. This is shown in the zoomed-in view of the detector in Fig.~\ref{fig:scheme}.

In a second EBL step, a pattern for normal metal leads (brown) was defined on top of the ALD-grown insulator. Importantly, a square-shaped normal metal quasiparticle (qp) trap (yellow) forms a considerable part of the counting node. Without the inclusion of the qp trap, the middle island would be formed only by the superconducting electrodes of the L and R turnstiles (and their tunnel-coupled normal shadows), resulting in poor quasiparticle thermalization and hence extra pumping errors~\cite{knowles12,heimes14}. We performed reference measurements on samples with an identical normal metal qp trap, but which contained a continuous overlapping superconducting island between the L and R turnstiles. This resulted in no significant difference compared to the geometry of Fig.~\ref{fig:scheme}. The qp trap as well as the N leads, starting approximately $10\mum$ away from the turnstile junctions and gradually widening into bonding pads, were metallized by $2\nm$ of Ti followed by $30\nm$ of AuPd.

The third and final lithography step, relying on a Ge-based hard mask~\cite{pekola13}, was used to define the turnstile islands and lead patterns for multi-angle shadow evaporation of the ultrasmall NIS tunnel junctions. Quickly widening S leads help enhance qp thermalization~\cite{knowles12}. The dry development process of the Ge mask, based on reactive ion etching in $\mr{CF}_{4}$ and $\mr{O}_{2}$ plasmas, further facilitates good contacts between the AuPd qp traps and the superconducting leads. To fabricate the typically sub-$30\nm\times 30\nm$ junctions for the L and R turnstiles as well as the slightly larger detector junctions, $15\nm$ of aluminium was first deposited at zero tilt angle, and immediately oxidized {\it in-situ} for a few minutes in an atmosphere of a few millibars of pure oxygen. After the initial Al layer we deposited $30\nm$ copper to form the detector island and junctions, evaporated at a tilt angle resulting in the downwards-shifted shadow copy of the structures evident in the insets of Fig.~\ref{fig:scheme}. Finally, (after a second, optional oxidation to reduce the junction transparency) another $50\nm$ thick layer of Cu was deposited with an upwards shift to form the L and R turnstile islands and junctions, thus completing the structure.

All measurements were performed in a dilution refrigerator at the base temperature close to $50\mk$. The chip was enclosed in an indium-sealed rf-tight sample holder~\cite{saira12}, with all the signal lines filtered by Thermocoax cables (shorter length for the L and R gates for higher bandwidth), and the sample stage was thermally anchored to the mixing chamber. The L and R gate signals were connected to two independent, synchronized outputs of a 2-channel arbitrary waveform generator, or to the outputs of two separate but synchronized 1-channel generators. After preamplification and analog low-pass filtering (cutoff at $50\khz$), the detector voltage was recorded by an optically isolated 16-bit PCI digitizer at the sampling rate $\fs=200\khz$ and saved to disk for further analysis and optional digital filtering.

\section{Device characterization}
\begin{figure}[!htb]
\includegraphics[width=\columnwidth]{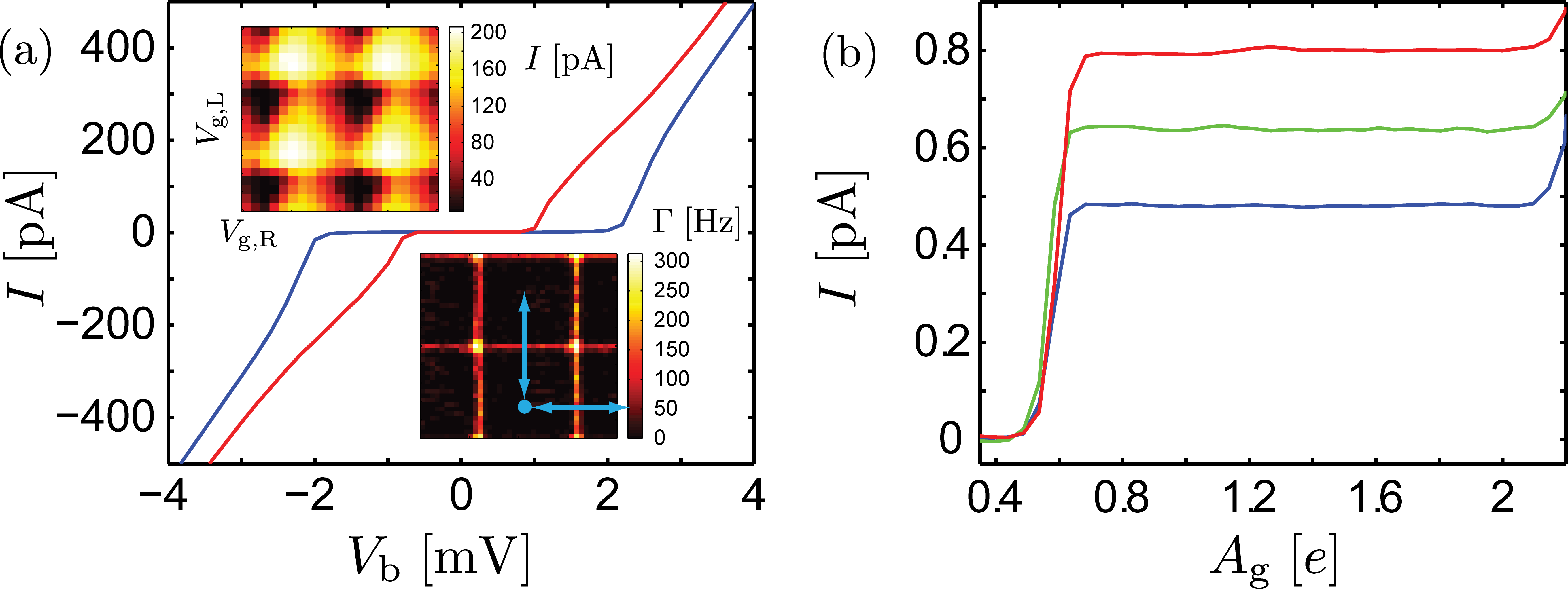}
\caption{(color online) \figta Current--voltage characteristic of the series turnstiles L and R. The two points at each $\vdut$ show the minimal and maximal current when the gate voltages $\vgl$ and $\vgr$ are swept over several periods. The top inset shows the gate modulation of the current $I$ at fixed $\vdut=2\mv$. The bottom inset displays the gate-dependent residual tunnel-event rate observed by the detector when the series turnstiles are biased at fixed $\vdut=-50\muv$ and the gate offsets are scanned. The light blue arrows sketch how the gates are operated sequentially during electronic pumping. \figtb From top to bottom: plateaus of the average pumped current $I$ as a function of the drive amplitude $\ag$ (peak-to-peak) under continuous pumping operation at 3, 4, and $5\mhz$, respectively.} \label{fig:dut}
\end{figure}

Figure~\ref{fig:dut}~\figta summarizes results of the series turnstile characterization in the absence of all time-dependent drive signals, prerequisite to measurements under pulsed driving to find proper operating conditions. First, the main panel of~\figta shows the current--voltage characteristics of the series devices. The blue (red) curve exhibiting maximum (minimum) blockade at low bias voltages $\vdut$ corresponds to both turnstiles gate-tuned to maximum Coulomb blockade, {\it i.e.}, integer charge state, or charge degeneracy point, respectively. The asymptotic slopes of this plot yield the total tunneling resistance $\rtot\approx10\Mohm$. The remaining gate modulation even at $\vdut>4\mv$ indicates a degree of asymmetry in the individual junction resistances, which can be expected for the ultrasmall contacts. The minimum voltage below which the current $\idut$ is strongly suppressed corresponds to $4\Delta/e\approx 900\muv$, typical for the series combination of two Al-based SINIS structures with $\Delta\approx 225\muev$. In contrast, the maximum blockade voltage $4\Delta+\esigma$ gives an estimate $\esigma\approx 1\mev$ for the sum of the charging energies of the individual turnstiles.

The top left inset of Fig.~\ref{fig:dut}~\figta displays a surface plot of the current $\idut$ through the turnstiles at fixed bias $\vdut=2\mv$ as a function of the two gate voltages: $\idut$ is $e$-periodic in both $\vgl$ and $\vgr$, and maximized when each turnstile is tuned to charge degeneracy. The lack of skewness, {\it i.e.}, distortion of the underlying square lattice of current maxima, indicates negligible coupling between the L and R gate signals. Comparing with the envelope curves in the main panel of Fig.~\ref{fig:dut}~\figa, also here the cross section shape suggests asymmetry of the turnstile junctions. Based on SEM observations, a difference of $50\%$ in the junction areas is typical.

At $\vdut\ll 1\mv$ the average current $\idut$ is strongly suppressed. However, in this bias regime the detector comes to play as a direct probe of the charge fluctuations through the turnstiles: The bottom right inset of Fig.~\ref{fig:dut}~\figta shows a 2D slice of the total rate $\Gamma$ of tunneling events onto or off the central island at $\vdut=-50\muv$. This plot inherits its shape from the behavior of $\idut$ as a function of the two gate voltages, evident as strong peaking of $\Gamma$ at double charge degeneracy. As $\vdut$ is increased, the areas of elevated $\Gamma$ around these points and the lines connecting them grow larger. Nevertheless, even at $\vdut\approx 400\muv$, essential for the driven operation of the turnstiles, a large range of suppressed $\Gamma$ remains around each gate offset corresponding to an integer charge state on the corresponding turnstile island. Notably, in the present experiment where the electrons are counted in the S lead, we are able to probe the tunneling rates even when the turnstile islands are in Coulomb blockade. This was not the case for the previous SINIS counting experiments~\cite{saira12,maisi11}, and constitutes an essential ingredient needed for detection of higher order tunneling processes in SINIS structures.

In Fig.~\ref{fig:dut}~\figtb we demonstrate that, under continuous drive, electrons can be transferred through the two series turnstiles as expected. The plot shows the average pumped current $\idut$ at $\vdut=400\muv$ when the L and R gates were driven simultaneously, with relative delay time $\tau=0$, by pulses of 50 \% duty cycle and increasing amplitude $\ag$ (peak-to-peak), centered around charge degeneracy points for each turnstile. The blue, green, and red curve from bottom to top correspond to continuous pumping at $\frep=3$, 4, and $5\mhz$, respectively. In this measurement, limited by the short averaging time, the resulting average currents of 0.48, 0.64, and $0.80\pa$ on the plateau are within 1 \% of the expected values given by $e\frep$.

\begin{figure}[!htb]
\includegraphics[width=\columnwidth]{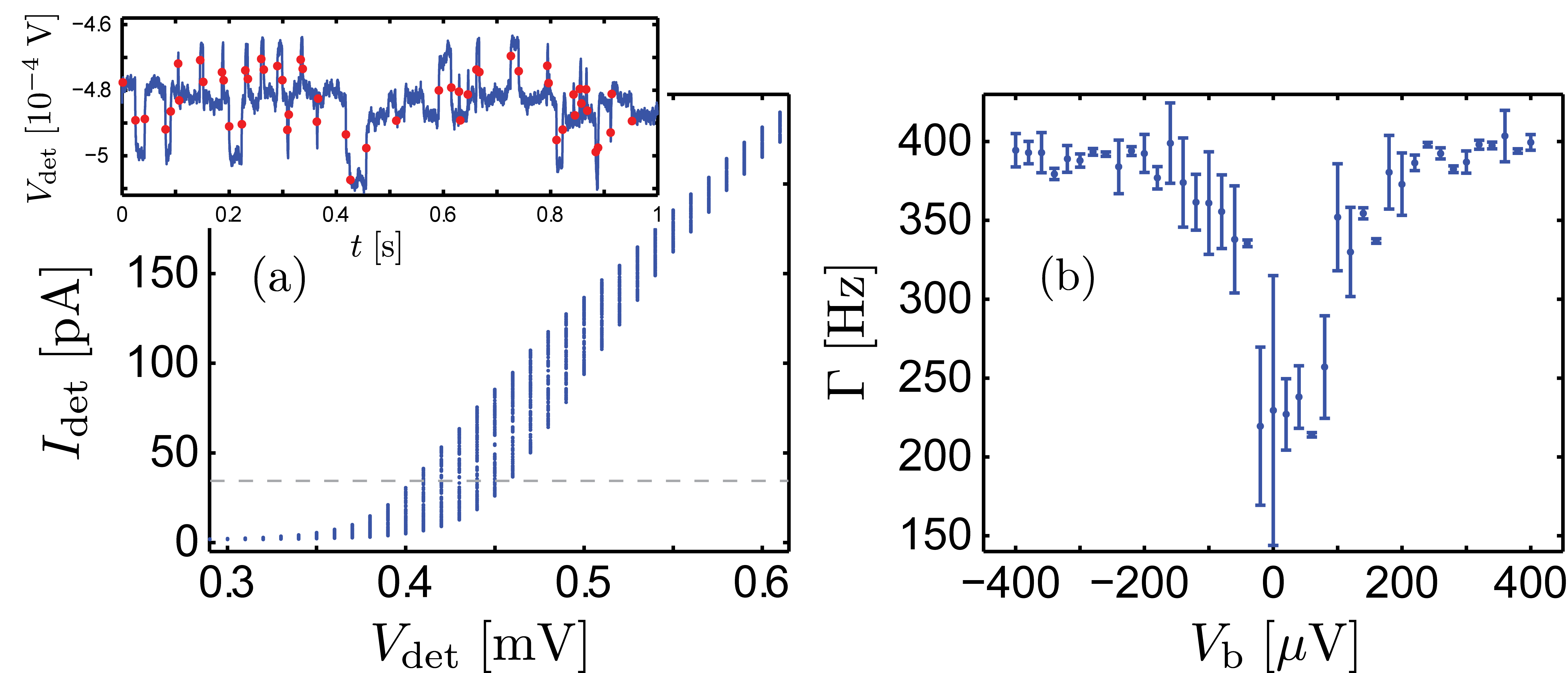}
\caption{(color online) \figta Detector current--voltage characteristic, with the points at each $\vdet$ obtained by sweeping the gate voltage $\vgdet$ over several periods. The horizontal dashed line indicates the fixed bias current used for most of the measurements. The inset shows a typical time trace of the detector signal in the absence of time-dependent drives of $\vgl$ and $\vgr$. The red dots indicate tunneling events identified by a simple edge-detecting algorithm. \figtb Bias voltage dependence of the maximum observed event rate $\Gamma$ of the undriven system, corresponding to both turnstiles tuned to charge degeneracy. The rates are extracted as the maxima of 2D gate scans similar to the bottom inset of Fig.~\ref{fig:dut}~\figa. The error bars show the standard deviation from a few repetitions of the measurement.} \label{fig:dc}
\end{figure}

The main panel of Fig.~\ref{fig:dc}~\figta displays the IV characteristic of the detector SET with $\rtdet\approx1\Mohm$. The dots at each $\vdet$ indicate the range of currents $\idet$ flowing through the detector as the gate voltage $\vgdet$ is swept over a period corresponding to several $e$. The bias current $\idet\approx 40\pa$, shown by the dashed gray horizontal line and employed throughout the electron counting experiments described in this work, was chosen based on high sensitivity and small backaction onto the turnstiles. The inset of Fig.~\ref{fig:dc}~\figta shows a typical time trace of the detector signal for an undriven system at $\vdut=-50\muv$, with $\vgl$ and $\vgr$ tuned close to charge degeneracy. Here we choose to operate the detector at fixed current (and gate voltage $\vgdet$) and record the varying voltage $\vdet$. To facilitate extraction of the rate $\Gamma$ from such traces, the red dots indicate edges identified by a simple algorithm based on threshold detection and numerical differentiation of the detector signal after digital low-pass filtering. We note that the detector could equally well be operated at a fixed bias voltage $\vdet$ and thus changes in $\idet$ would be monitored. For the present measurements current bias was preferred due to higher signal-to-noise ratio.

In Fig.~\ref{fig:dc}~\figtb we plot the $\vdut$-dependence of the maximum of the event rate $\Gamma$, corresponding to both turnstiles at charge degeneracy. In this sample we found the voltage applied to the M-gate to have no clearly distinguishable effect on the rates, and this gate electrode coupled to the middle island was kept grounded during the majority of the measurements. For pumping operation under pulsed drive, described in the following sections, it is important to note that in general in our measurements only one of the turnstile gates was driven at any given time of the operation cycle, {\it cf.} the blue arrows in the bottom inset of Fig.~\ref{fig:dut}~\figa.

\section{Detection of all electrons at slow repetition rates}
\begin{figure}[!htb]
\includegraphics[width=\columnwidth]{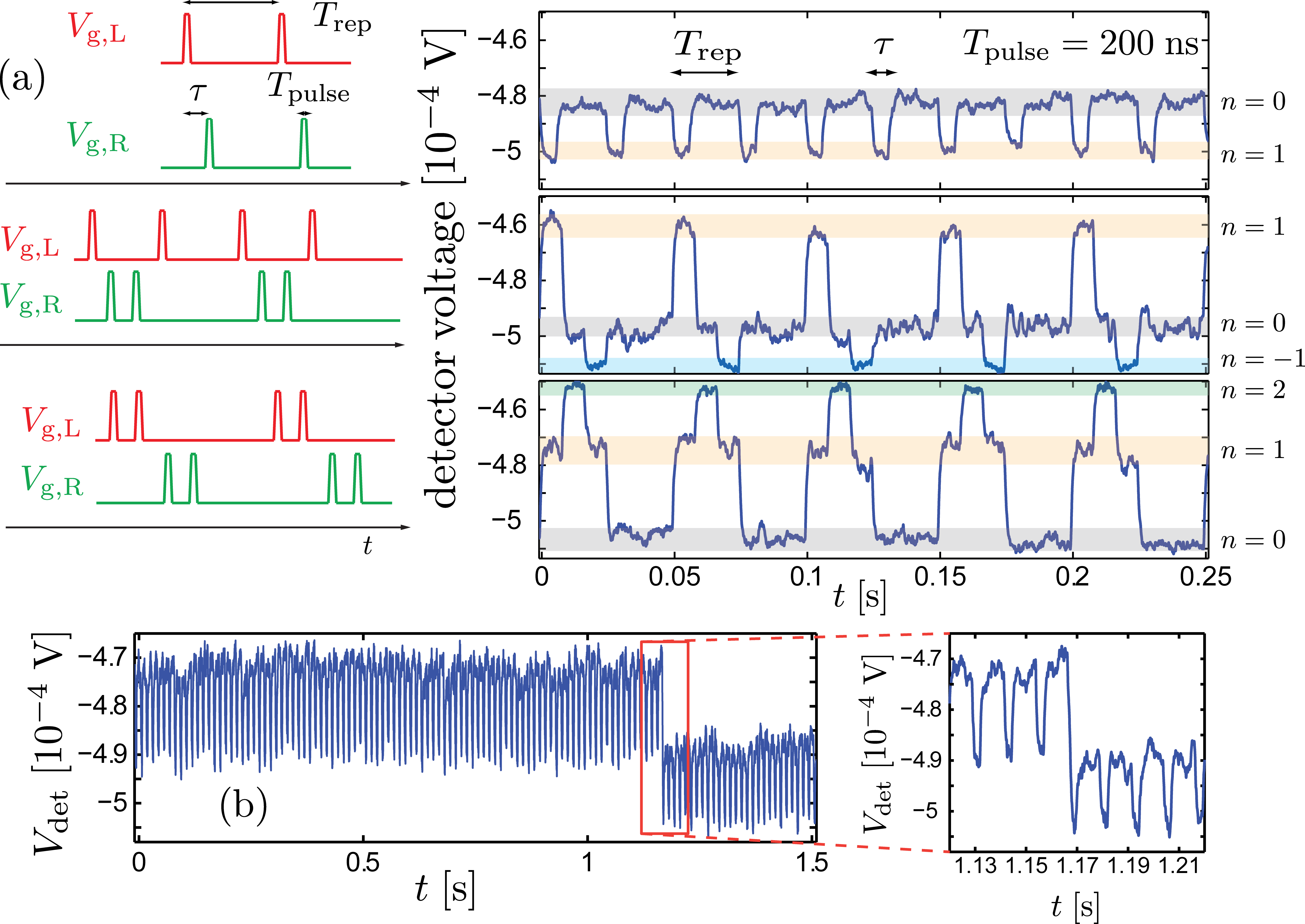}
\caption{(color online) \figta Examples of control sequences and corresponding detector signals during slow manipulation of the charge state on the counting node (see the main text for details). \figtb Detection of each electron over a longer time span of slow sequential pumping operation. An error event where turnstile L transfers two electrons occurs around 1.2 s. Such relatively rare errors remain distinguishable also with a bandwidth-limited detector even at high $\frep$ when each transferred charge is no longer resolvable. In this limit, however, the setup with a single counting node cannot discriminate whether the error is due to missed or extra electron tunneling.} \label{fig:pump}
\end{figure}

We now consider sequential operation under drive with low duty cycle and repetition frequency. The main finding of the present work is illustrated in Fig.~\ref{fig:pump}~\figa: The L and R turnstile gates are driven sequentially by the pulse sequences sketched on the left, where each of the gate voltage pulses of amplitude close to $1e$ is expected to ideally result in the transfer of one electron through the turnstile in question. As evidenced by the time traces of the detector voltage on the right, this is indeed what we directly observe -- the detector signal and hence the charge state on the middle island depends only on the repetition frequency $\frep$ and the phases (relative time delay $\tau$) of the two drive signals, whereas no dependence on the length $\tpulse\ll\fs^{-1},\frep^{-1},\tau$ of the individual drive pulses is seen. The three example sequences correspond to the transfer of $+1\rightarrow -1$ (top), $+1\rightarrow -1 \rightarrow -1 \rightarrow +1$ (middle), and $+1\rightarrow +1 \rightarrow -1 \rightarrow -1$ (bottom) electrons to the middle island. Each timing diagram sketches two identical cycles of the drive, with each cycle followed by a slight delay for easier visual recognition. The colored bands and the numbers on the right indicate the number of electrons on the counting node, relative to the value at the start of the drive.

The strong detector coupling is evident in the detector traces in Fig.~\ref{fig:pump}~\figa: Notably, the vertical scale is the same in all three panels, illustrating that the nonlinearity of the detector gate modulation becomes relevant after the charge state on the middle island changes by only a few electrons. Note also that the detector is operating on a different slope of its gate modulation in the top panel compared to the other two sequences -- in this case pulsing $\vgl$ results in a step down in the detected signal.

In Fig.~\ref{fig:pump}~\figta we considered different sequences of manipulation of the charge state on the counting node, for the duration of a few cycles of the periodic drive pulse trains. Figure~\ref{fig:pump}~\figtb extends this with a typical time trace over a longer span, demonstrating the sequential transfer of more than 100 electrons through the counting node without errors. Here, $\vdut$ was set to $-400\muv$ and pulses of $500\ns$ length and $1e$ amplitude were applied to $\vgl$ and $\vgr$ at $\frep=80\hz$ with $\tau=\trep/4$. The typical observed duration of faultless operation varied from 0.5 up to 10 seconds, in line with the residual event rate [{\it cf}. Fig.~\ref{fig:dut}~\figa] at the operating points when no drive pulses are applied. Also evident in Fig.~\ref{fig:pump}~\figtb is one error event where an extra electron is transferred through turnstile L. In such measurements at low repetition rates $\frep$ up to $100-200\hz$, where the timing of each tunneling event is clearly resolved by the detector, in spirit of Ref.~\onlinecite{wulf13} we can reliably identify in which of the two turnstiles the errors appear to originate from, the direction in which they occur, or whether they are caused by background charge jumps. In Ref.~\onlinecite{fricke14} the advantage of two detectors is that such information is available even at faster repetition rates. We aim to test this in future experiments. Moreover, based on straightforward modeling, it is further possible to estimate, e.g., the probability of misattribution of one missed tunneling in turnstile L to one extra electron tunneling in turnstile R, which would result in close to identical detector signals.

The operation points for the measurements in Fig.~\ref{fig:pump} were determined by first scanning the L gate offset while the R offset was kept fixed at some constant value. Under pulsed, constant-amplitude drive at 50 \% duty cycle the center of the resulting plateau in the average pumped current was then determined. The L gate offset was set to this optimal value, and another scan was subsequently performed but this time by varying the R offset. Such a gate offset search method based on the average pumped current under continuous drive was found to reliable as it is possible to keep $\vdut$ constant once set to the desired value. The advantage here is that the only action needed to transition between the the faster continuous pumping measurements in Fig.~\ref{fig:dut}~\figtb and the representative on-chip electrometer traces in Fig.~\ref{fig:pump}~\figta is a change in the duty cycle and repetition frequency of the L and R gate drives.

\section{Detection of pumping errors only at higher repetition rates}
\begin{figure}[!htb]
\includegraphics[width=\columnwidth]{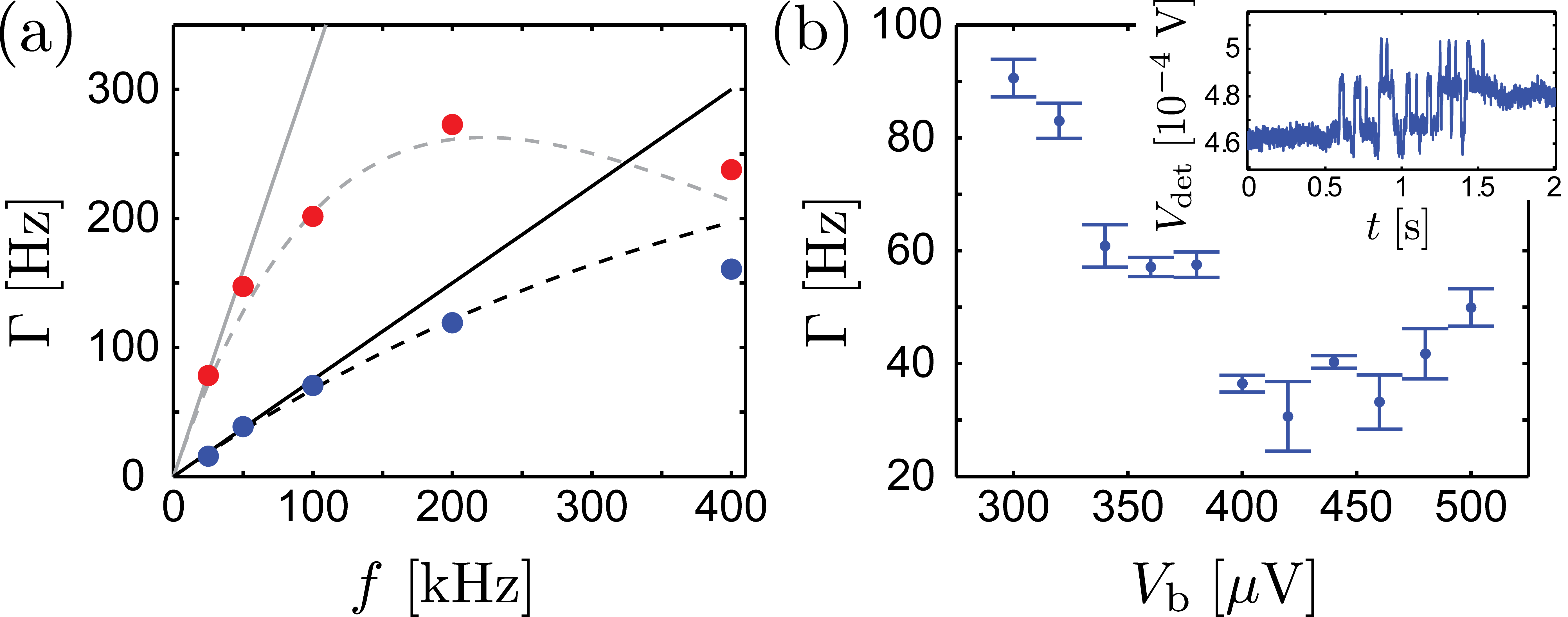}
\caption{(color online) \figta Observed error rate $\Gamma$ as a function of the drive frequency $\frep$, at $\vdut=-300\muv$ (red/light symbols), and at $\vdut=-420\muv$ (blue/dark). The respective gray and black solid lines are straight lines through origin, showing the expected linear behavior. The dashed lines include the effect of limited detector bandwidth. \figtb Bias voltage dependence of the error rate at the fixed repetition frequency $\frep=50\khz$. The inset shows a typical time trace of error events recorded at $\vdut=400\muv$, $\frep=50\khz$, and $\tpulse=100\ns$, with the pulsed drive switched on/off every 1 s. Here, the drive is on for approximately $0.5\s<t<1.5\s$.} \label{fig:fdep}
\end{figure}

We next describe detection of pumping errors at drive frequencies $\frep$ too high to distinguish each electron entering or exiting the counting node, but on the other hand low enough that the error rate remains in the sub-kHz bandwidth of the detector. For our present device, this limits the highest usable $\frep$ to around $100\khz$ or less. The symbols in Fig.~\ref{fig:fdep}~\figta show error rates $\Gamma$ extracted from 30 s long time traces of the detector signal. Here, the pulsed turnstile drive with $\tpulse=100\ns$ and $\tau=\trep/4$ at repetition frequencies $\frep$ up to $400\khz$ was switched on/off at the rate of $1\hz$ to help verifying the stability of the gate offsets. The trace in the inset of Fig.~\ref{fig:fdep}~\figb, obtained at $\vdut=400\muv$ and $\frep=50\khz$, illustrates a typical error signal. During the `drive off'-sections the detector registers a small number of extra counts due to the background rate, {\it cf.} Fig.~\ref{fig:dc}, but this contribution remains negligible compared to jumps recorded during the driven operation. The red and blue symbols correspond to measurements at $\vdut=-300\muv$ and $-420\muv$, respectively. Assuming independent $1e$ error events, we expect $\Gamma$ to scale linearly with $\frep$. The black and gray solid lines plot $\Gamma_0=r\frep$ with the respective relative error rates $r=3.2\times 10^{-3}$ and $7.5\times 10^{-4}$, in reasonable agreement with the detected rates at $\frep\lesssim 100\khz$.

At higher $\frep$ the measured values deviate increasingly from the expected linear behavior, and for $\vdut=-300\muv$ display even a tendency to decrease. We attribute this to the limited detector bandwidth: When two events happen within or almost within the $\approx 1\ms$ rise time, they cannot be reliably distinguished by the edge detection algorithm. The dashed lines model this by an effective error rate $\gammaeff=\Gamma_0\exp(-\Gamma_0\taudet)$, {\it i.e.}, the `intrinsic' rate multiplied by the estimated probability of two independent events with interval larger than the detector time constant $\taudet\approx 1.4\ms$. The initial, close to linear increase of the observed $\Gamma$ with $\frep$ demonstrates we are able to reliably count the number of pumping errors. However, compared to Ref.~\onlinecite{fricke14} with 3 pumps in series, monitored by two independent detectors on two separate counting nodes, we are unable to tell the direction of the errors or in which turnstile they occurred. In contrast to Fig.~\ref{fig:pump}~\figta where $\frep\ll\fs$ we can no longer deduce this based on detailed timing information of individual tunneling events.

In Fig.~\ref{fig:fdep}~\figtb we display a typical $\vdut$-dependence of the error rate, here obtained at the fixed value $\frep=50\khz$. The origin of this reproducible variation of the error rate on $\vdut$ needs further investigation. Although qualitatively in line with arguments~\cite{pekola08,averin08} based on balance between missed tunneling and extra unwanted tunneling and leading to the optimum bias voltage $\vdut\approx\Delta/e$ per turnstile, we find only a weak dependence on the pulse length $\tpulse$. The effect of $\vdut$ on the detector gate position was not directly compensated in this measurement. However, we checked that detector sensitivity did not vary significantly due to changes in $\vdut$. Also, the as-expected initial linear scaling of the error rates with $\frep$ suggests this is not the likely cause. In addition, detector backaction~\cite{lotkhov12}, at least not due to simple heating, appears not to be a likely origin of the effect: for the non-driven rate measurements in Fig.~\ref{fig:dc} we checked that reducing $\idet$ to $20\pa$ did not significantly influence the rates, it only reduced the signal-to-noise ratio.

\section{Discussion and perspectives for future work}
To transform the on-chip electrometry from estimation of the total number of errors events only into actual error accounting and self-referenced operation, in future work we will look into extending the scheme of the present experiments by adding a third turnstile and a second detector in series with the existing setup. We can thus regain information about the direction of the errors, even for $\frep\gtrsim\fs$, via correlation analysis of the two detector signals~\cite{fricke14}. What is more, to increase the bandwidth of the detector, it can be operated in an RF-SET configuration using an on-chip resonant circuit for impedance matching~\cite{schoelkopf98,bylander05}, or connected directly to a cryogenic preamplifier, located either at a higher temperature stage of the dilution refrigerator~\cite{vink07} or directly at the mixing chamber stage~\cite{curry15}. Likewise, feedback control may be implemented for the detector and M gate voltages $\vgdet$ and $\vgm$, respectively.

More generally, looking at practical requirements for the turnstile--detector combination to perform reliable error counting, we can immediately write down simple criteria: For a slow detector capable of resolving pumping errors occurring at $500\hz$, a pumped current of $50\pa$ requires $\frep\approx 300\mhz$ and sets a strict limit of $2\times10^{-6}$ for the relative error rate of the turnstiles. In contrast, for a fast detector capable of coping with a $250\khz$ event rate, already $9\times10^{-4}$ is sufficient, well within reach by the present turnstiles. Moreover, if parallel operation of the turnstiles~\cite{maisi09} can be used to bring the current requirement down to, say, $10\pa$, the relative error rates are further relaxed by the same factor.

Besides developing the in-situ charge counting, we are currently investigating ways to improve the intrinsic turnstile accuracy. The principle of charge monitoring between two devices also allows us to study fundamental error mechanisms in hybrid turnstiles, the probing of which has not been possible in earlier setups. Direct measurement of the cutoff frequency for missed tunneling constitutes one example. Another possibility is real-time observation of Cooper pair -- electron cotunneling (CPE)~\cite{averin08}, a higher order process where effectively the tunneling of one electron through one of the turnstile junctions occurs coherently with the tunneling of another electron through both junctions. This takes place as two-electron Andreev reflection (AR) in one junction combined with simultaneous one-electron process in the other junction.

Two-electron Andreev tunneling events have been observed in a setup where the charge state on the N island of an individual SINIS turnstile was monitored in real time by a SET electrometer~\cite{maisi11}. Such a configuration, however, cannot directly discriminate CPE events from regular one-electron tunneling. In contrast, in the scheme of Fig.~\ref{fig:scheme} where the superconducting counting node is now monitored instead, CPE and AR, both changing the charge state by $2e$, can be distinguished by their differing dependencies on the bias voltage $\vdut$. Furthermore, in a setup with at least three series turnstiles and two monitored counting nodes, these tunneling events can be directly labeled based on the detector signals. For investigating the CPE process, it is advantageous to have a Coulomb blockaded island. In the configuration of Ref.~\onlinecite{maisi11} the counting of the charges on the N island is not possible at such gate offsets because of the fast tunneling rate back to the lowest electron number state. Thus performing the counting 'at the lead' as demonstrated here is really an asset for the present experiment. In principle CPE could be investigated at degeneracy condition as well, but in practice there is always a residual tunnel rate remaining which hinders it.

Finally, in view of recent pumping results with NISIN turnstiles operated in finite magnetic fields~\cite{taupin15}, it would be interesting to perform similar counting experiments with a sample where the roles of N islands and S leads are reversed -- there would be no need for a separate qp trap or N leads fabricated in an extra step.

\section{Conclusions}
To summarize, we have experimentally investigated the feasibility of real-time error detection for SINIS turnstiles. Limited by the bandwidth of the dc SET detector, we have demonstrated proof-of-principle monitoring of the charge pumped through an island between two turnstiles. With metallic tunnel junction-based devices it is straightforward to integrate the detector and the pumps to the same sample and to arrange a sufficient coupling between them. The error rates of the individual turnstiles are currently mostly limited by slow quasiparticle relaxation, {\it i.e.}, overheating in the superconducting leads, in particular in the current configuration of a counting node.

\section{Acknowledgments}
We thank T. Faivre for assistance with sample fabrication, A. Kemppinen, J. Lehtinen, and E. Mykk\"anen for help with reference measurements, and V. Kashcheyevs, J. V. Koski, A. Manninen, M. Meschke, Y. Pashkin, O.-P. Saira, and A. Zorin for helpful discussions. Initial part of the work was supported by the EMRP project "Quantum Ampere". J. T. P. acknowledges support from Academy of Finland (Contract No. 275167).

\end{document}